\documentclass[a4paper]{report}
\usepackage[utf8]{inputenc}
\usepackage[T1]{fontenc}
\usepackage{RJournal}
\usepackage{amsmath,amssymb,array}
\usepackage{booktabs}

\usepackage{pmboxdraw} 

\begin{document}

\sectionhead{Contributed research article}
\volume{XX}
\volnumber{YY}
\year{20ZZ}
\month{AAAA}

\begin{article}
\title{Writing R Extensions in Rust}
\author{by David B.\ Dahl}

\maketitle

\abstract{This paper complements ``Writing R Extensions,'' the official guide
for writing R extensions, for those interested in developing R packages using
Rust. It highlights idiosyncrasies of R and Rust that must be addressed by any
integration and describes how to develop Rust-based packages which comply with
the CRAN Repository Policy.  This paper
introduces the \pkg{cargo} framework, a transparent Rust-based API which wraps
commonly-used parts of R's API with minimal overhead and allows a programmer to
easily add additional wrappers.}

\section{Introduction}

Computationally-intensive R packages are typically implemented using C, Fortran,
or C++ for the sake of performance. The R Core Team maintains a document called
``Writing R Extensions'' which describes R's API for creating packages.
This paper supplements that official guide by (i) discussing issues
involved in the integration of R and Rust and (ii) providing an R package to help those
interested in writing R packages based on Rust.

While both R and Rust provide
foreign function interfaces (FFI) based on C \citep{C}, each language has its
own idiosyncrasies that require some care when interfacing with the other. Packages published on CRAN (\url{https://cran.r-project.org/}) are
subject to the CRAN Repository Policy. This paper also describes how to avoid pitfalls which may prevent acceptance of a Rust-based
package on CRAN or waste time of CRAN maintainers and package contributors.

The paper introduces the newly-released \CRANpkg{cargo} \citep{cargo}
package which provides a framework for developing CRAN-compliant R packages
using Rust and shows how to make Rust-based wrappers for R's C API. It is
important to emphasize that a package developed with the \pkg{cargo} framework
does \emph{not} depend on the \pkg{cargo} package, either in source or binary
form. That is, the \pkg{cargo} package produces an R package structure with all
the necessary Rust code and scripts such that the package is then independent of
the \pkg{cargo} package.  Developers can then extend the framework for their own
purposes within the generated package structure. Further, although a source
package will obviously depend on Rust, there are no runtime dependencies on
Rust or any other libraries, resulting in a binary package that is easy
for others to use.
Separate from package development, the \pkg{cargo} package also allows Rust code to be directly embedded in an R script.

One of the purposes of this paper is to encourage developers to consider Rust
for writing high-performance R packages. A second purpose
is to discuss technical issues which arise when interfacing R and Rust and to
document the design choices of the \pkg{cargo} framework. The \pkg{cargo} framework seeks to: (i) provide a Rust interface
for commonly used parts of the R API, (ii) show the developer how they can
easily extend the framework to cover other parts of the R API, (iii) minimize
the runtime overhead when interfacing between R and Rust, and (iv) be as
transparent as possible on how the framework interfaces R and Rust.  This paper
assumes some familiarity with ``Writing R Extensions'' and package
development using R's API.
The paper also assumes some familiarity with Rust.  The interested reader is
directed to a plethora of resources online, including ``The Rust Programming
Language'' (\url{https://doc.rust-lang.org/stable/book/}).

The paper is organized as follows. A brief history on Rust and its use in R
is outlined.
Setting up the Rust toolchain for R package
development is discussed next, followed by an overview of the various parts of
an R package using the \pkg{cargo} framework.  Low-level and high-level
interfaces between R and Rust are introduced.  Threading issues and
seeding a random number generator are also discussed. Defining a R function by
embedding Rust code directly in an R script is shown.  Finally, the paper ends
with benchmarks and concluding comments.

\section{Background on Rust and its use in R}

Rust (\url{https://www.rust-lang.org/}) is a statically-typed, general-purpose
programming language which emphasizes memory safety without compromising runtime
performance.  Its memory safety guarantees (against, e.g., buffer overflows,
dangling pointers, and race conditions) are achieved through the language's
design and the compiler's borrow checker.  This avoids the memory and CPU
overhead inherent in garbage-collected languages.  Concurrent programming is
straightforward in Rust, where most concurrency errors are compile-time errors
rather than difficult-to-reproduce runtime errors. Developer productivity is
aided by rustup (toolchain installer and upgrader), Cargo (package manager for
downloading dependencies, publishing code, and building dependencies and code),
Rustfmt (automatic code formatter), and Clippy (linting tool to catch common
mistakes and improve performance and readability).

Rust first appeared in 2010 as a Mozilla project, had its first stable release
in 2015, and has been rated the ``most loved programming language'' in the Stack
Overflow Annual Developer Survey
(\url{https://insights.stackoverflow.com/survey/}) every year since 2016. The
Rust Foundation was formed in 2021 with the founding members Amazon Web Services,
Google, Huawei, Microsoft, and Mozilla. Google recently announced support for
Rust within Android Open Source Project (AOSP) as an alternative to C and C++.
Experimental Rust support for developing subsystems and drivers for the Linux
kernel has been submitted.  Linus Torvalds has been quoted on several occasions
as being welcoming of the possibility of using Rust alongside C for kernel
development.

Members of the R community have also been interested in Rust.  The first major
effort to integrate R and Rust appears to have started in early 2016 with the
now-deprecated \url{https://github.com/rustr} project. The first Rust-based
package appeared on CRAN in 2018 with Jeroen Ooms' \CRANpkg{gifski} package
\citep{gifski}, with an accompanying presentation at the 2018 European R Users
Meeting (eRum2018) describing how a developer can use Rust code in an R package.
The approach requires the package developer to write C code which then calls
Rust code.  Under this approach, the Rust code itself does not have access to
R's API.

In 2019, my \CRANpkg{salso} package \citep{salso} was the second Rust-based
package on CRAN. It followed \pkg{gifski}'s approach of writing C code that
calls Rust code. Around the time that the third Rust-based package
\CRANpkg{baseflow} \citep{baseflow} was accepted to CRAN, the CRAN maintainers
noted that \pkg{gifski}, \pkg{salso}, and \pkg{baseflow} violated the policy
that ``packages should not write ... on the file system apart from the R
session’s temporary directory'' since Cargo caches downloaded dependencies by
default and uses all available CPU cores.  This inspired me in early 2021 to
write the \pkg{cargo} package to facilitate using Cargo in conformance with
CRAN's policies and to download precompiled static libraries in case the
required version of the Rust toolchain is not available on a particular CRAN
build machine. It also became clear that writing C code that glues the R and
Rust code is tedious, error prone, and difficult to refactor.  As such, I
expanded the \pkg{cargo} package to facilitate developing Rust-based packages
that avoid the need to write the C glue code, allowing R to call directly into
Rust code and allowing Rust code to callback into R's API directly. In 2021,
CRAN accepted \CRANpkg{caviarpd} \citep{caviarpd} as another package developed
using the \pkg{cargo} framework and the \pkg{salso} package was ported to the
framework.

Another exciting project that interfaces R and Rust is the extendr
project (\url{https://github.com/extendr}). Andy Thomason started working on the
extendr project in 2020, attracting Claus Wilke and several other developers.
The extendr project seeks not
only to facilitate writing R packages in Rust, but also to embed the R
interpreter in a Rust program. In 2021, the project released the
\CRANpkg{rextendr} package \citep{rextendr} on CRAN to facilitate developing
Rust-based packages. In 2021, the \pkg{baseflow} package was ported to use
\pkg{rextendr}, and the \CRANpkg{string2path} package \citep{stringpath} became
another package on CRAN developed with the aid of \pkg{rextendr}.

Those interested in interfacing Rust and R should keep an eye on the extendr project as it continues to evolve.
The project is working
to provide extensive automatic conversion between R types (e.g., vectors,
lists, data.frames, environments, etc.) and Rust types, including attempts to
handle thorny issues such as R's missing value \code{NA} and R's fluidity in vectors
of storage mode \samp{double} and \samp{integer}. It aspires to eventually provide a Rust interface for all
of the functionality provided by the R API to alleviate the Rust developer from
having to dive into the details of R's API.

The \pkg{rextendr} package and the \pkg{cargo} package both seek to provide functionality to develop R packages
which can call directly into Rust and call back to R from Rust.
The extendr project is hosted on public GitHub repositories and is under
rapid development; their project shows what is possible and their open discussion influenced some of my choices for the \pkg{cargo} package.
The \pkg{rextendr} package and the \pkg{cargo} package address various technical issues
differently and choose different design trade-offs.
The advantage of the \pkg{cargo} package is its transparency and extendability,
whereas the benefit of the \pkg{rextendr} package is that it provides many
behind-the-scenes type conversions and aims to be more comprehensive. The lean nature of the
\pkg{cargo} framework makes it simple to understand how R and Rust interface.

\section{Installing the Rust toolchain}

Developing Rust-based R packages requires the installation of several tools.
The first step is to install the usual
toolchain bundle to compile C/C++/Fortran packages for the chosen operating system.
For example, on Windows, install Rtools
(\url{https://cran.r-project.org/bin/windows/Rtools/}).  On MacOS, follow the
instructions here: \url{https://mac.r-project.org/tools/}.
Next, install the Rust toolchain;
the recommended way is rustup
(\url{https://rustup.rs}). On MacOS and Linux, this entails executing a single
shell command. There are a few more steps on Windows.  One must download
rustup‑init.exe from \url{https://rustup.rs}, execute \samp{rustup-init.exe -y
--default-host x86\_64-pc-windows-gnu}, and then (in a new terminal so the
changes to the PATH are picked up), execute \samp{rustup target add
i686-pc-windows-gnu}.

Rust has a six-week release cycle and the Rust toolchain is easily upgraded with
\samp{rustup update}. Incompatible changes are opt-in only, so new releases are
always guaranteed to run old code.  Because there are immediate benefits and no
costs to upgrading, Rust developers frequently develop against the latest Rust
version to take advantage of new features, optimizations, and bug fixes.

Rust's rapid release cycle, however, presents challenges when submitting
Rust-based packages to CRAN, as a CRAN build server may not have a recent
version of the Rust toolchain.  Moreover, the toolchain may not even be
available on a particular CRAN
build machine. A solution to this problem is
to host precompiled static libraries, which
is a common practice for packages unrelated to Rust. (See, for
example, \url{https://github.com/rwinlib}.)
The \pkg{cargo} framework provides tools to download a precompiled static library for the Rust component of an R package when a sufficient version of the Rust toolchain
cannot be found during installation.  Compiling from source is completely
supported by the \pkg{cargo} framework if the Rust toolchain meets the minimum
version specified in the package's \file{DESCRIPTION} file.

\section{Overview of package development using the \pkg{cargo} framework}

The \pkg{cargo} package facilitates the development of R packages based on Rust.
Install \pkg{cargo} from CRAN using \code{install.packages("cargo")}.
 Development starts by creating a new package using, for example,
\code{cargo::new\_package("/path/to/package/foo")} to generate the package
\pkg{foo} at the filesystem path \file{/path/to/package/}. If using RStudio,
this can be accomplished using "File" -> "New Project..." -> "New Directory" ->
"R Package Using Rust and the 'cargo' Framework".  This generates and installs a
complete working package that developers can modify for their own needs.

The directory structure of the new \pkg{foo} package is:
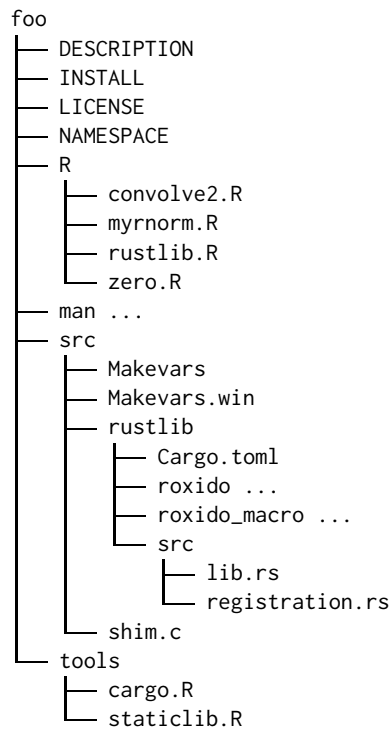
\begin{example}
        foo
        ├── DESCRIPTION
        ├── INSTALL
        ├── LICENSE
        ├── NAMESPACE
        ├── R
        │   ├── convolve2.R
        │   ├── myrnorm.R
        │   ├── rustlib.R
        │   └── zero.R
        ├── man ...
        ├── src
        │   ├── Makevars
        │   ├── Makevars.win
        │   ├── rustlib
        │   │   ├── Cargo.toml
        │   │   ├── roxido ...
        │   │   ├── roxido_macro ...
        │   │   └── src
        │   │       ├── lib.rs
        │   │       └── registration.rs
        │   └── shim.c
        └── tools
            ├── cargo.R
            └── staticlib.R
\end{example}
Several of the resulting files and directories are specific to packages
developed with the \pkg{cargo} framework. The \file{src/Makevars} and
\file{src/Makevars.win} direct R to use the \file{tools/staticlib.R} script to
compile the static Rust library defined in \file{src/rustlib} or, as a fallback,
to download a precompiled static library.  The download URL needs to be
provided in the \file{tools/staticlib.R} script. Notice that the
\file{DESCRIPTION} file has an entry \samp{SystemRequirements: Cargo (>= 1.54)
for installation from sources: see INSTALL file}.  The minimum required Cargo
version should be updated and the developer can determine this using cargo-msrv
(\url{https://crates.io/crates/cargo-msrv}). The \file{tools/cargo.R} script
finds and runs the Cargo package manager according to CRAN policies by, for
example, using no more than two CPU threads and downloading dependencies to a
temporary directory.  Unfortunately, the dependencies must then be redownloaded
and recompiled every time the package is reinstalled, which is a hassle during package
development. To avoid these limitations on a local development machine, the
package developer can add the followings to their personal \file{.Rprofile} file:
\begin{example}
        Sys.setenv(R_CARGO_SAVE_CACHE="TRUE")
        Sys.setenv(R_CARGO_BUILD_JOBS="0")
\end{example}
As an aid to other Rust-based packages \emph{not} using the \pkg{cargo} framework, the
functionality provided by the \file{tools/cargo.R} script is also available as
the \code{run} function in the \pkg{cargo} package.

There are several calls to the \code{.Call} function among the
scripts in the \file{R} directory.  The function in \file{R/myrnorm.R}, for
example, has \code{.Call(.myrnorm, n, mean, sd)} which calls the Rust function
\code{myrnorm} defined in \file{src/rustlib/src/lib.rs}:
\begin{example}[numbers=left,xleftmargin=6.2mm]
    mod registration;
    use roxido::*;

    #[roxido]
    fn myrnorm(n: Rval, mean: Rval, sd: Rval) -> Rval {
        unsafe {
            use rbindings::*;
            use std::convert::TryFrom;
            let (mean, sd) = (Rf_asReal(mean.0), Rf_asReal(sd.0));
            let length = isize::try_from(Rf_asInteger(n.0)).unwrap();
            let vec = Rf_protect(Rf_allocVector(REALSXP, length));
            let slice = Rval(vec).slice_double().unwrap();
            GetRNGstate();
            for x in slice { *x = Rf_rnorm(mean, sd); }
            PutRNGstate();
            Rf_unprotect(1);
            Rval(vec)
        }
    }
\end{example}
Notice that the \code{myrnorm} function has the \code{\#[roxido]} attribute and
takes three arguments \code{n}, \code{mean}, and \code{sd}, all of type
\code{Rval}, and returns a value of type \code{Rval}. The \code{\#[roxido]}
attribute is a procedural macro defined in
\file{src/rustlib/roxido\_macro/src/lib.rs} which adds the qualifiers
\code{\#[no\_mangle] extern "C"} when compiling to tell the Rust compiler to
make the \code{myrnorm} function callable directly from R. The attribute also
ensures that all arguments are of type \code{Rval} and that the return type is
\code{Rval}.  The \code{\#[roxido]} attribute also wraps the body of the
function in a call to Rust's \code{std::panic::catch\_unwind} since unwinding
from Rust code into foreign code is undefined behavior and likely crashes R.
When a panic is caught, it is turned into an R error which gives the
corresponding message from Rust and the line number of the panic.  The package
developer is encouraged to study the definition of the \code{\#[roxido]}
attribute in \file{src/rustlib/roxido\_macro/src/lib.rs} to better understand
the interface between R and Rust.

When a developer wants to make another Rust function callable by R, say a
function named \code{bar} taking two arguments \code{x} and \code{y}, the
developer adds the \code{.Call(.bar, x, y)} in a script under the \file{R}
directory of the package and then runs
\code{cargo::register\_calls("/path/to/package/foo")}.  This automatically
regenerates the \file{src/rustlib/src/registration.rs} file and does two things.
First, the updated file provides a stub for a
Rust function \code{bar} with arguments \code{x} and \code{y} in a commented-out block.  This stub
can then be copied to the \file{src/rustlib/src/lib.rs} file and the function
can be implemented. Second, code is generated to register functions when R loads
the shared library.  Again, the package developer is encouraged to study the
\file{src/rustlib/src/registration.rs} to see examples of how to call R's API
functions from Rust.

\section{Low-level interface to R's API}

The \code{myrnorm} function in Rust illustrates directly using R's API in Rust.
Line 7 of the listing is \samp{use rbindings::*}, which provides direct
access to R's API through Rust bindings.  These are automatically generated by
the bindgen utility (\url{https://rust-lang.github.io/rust-bindgen/}) from the
following R header files: \file{Rversion.h}, \file{R.h}, \file{Rinternals.h},
\file{Rinterface.h}, \file{R\_ext/Rdynload.h}, and \file{Rmath.h}, although only
those definitions and functions that are documented to be part of R's API (as
specific by ``Writing R Extensions'') should be used. The documentation for the
Rust bindings can be browsed by executing \samp{cargo doc --open} when in the
\file{src/rustlib/roxido} directory.  Note that most of the functions in the
\code{rbindings} module require an \code{SEXP} value, i.e., a pointer to R's
internal \code{SEXPREC} structure.  The \code{Rval} is defined as \samp{pub
struct Rval(pub SEXP)}, a \dfn{newtype} pattern that wraps the \code{SEXP}
value.  The newtype pattern provide type safety and encapsulation, which we
utilize in the high-level interface described in the next section. Because of
zero-cost abstraction, the Rust compiler generates code as if \code{SEXP} were
used directly.  The upshot is that, when calling R API functions, the
\code{SEXP} must be extracted from an \code{Rval} value, e.g., if \code{mean} is
an \code{Rval}, use \code{mean.0} to extract its \code{SEXP}, as in line 9.
Conversely, when returning from a function marked with \code{\#[roxido]}
attribute, wrap the \code{SEXP} value \code{x} in \code{Rval(x)}, as in line 17.

When accessing an R API function from Rust, care should be taken so that the R
function does not throw an error.  If Rust code calls an R function that throws
an error, a long jump occurs over Rust stack frames, which prevents Rust from
doing its usual freeing of heap allocations, resulting in a memory leak.  For
example, before calling \code{REAL(x)} to receive a pointer of type \code{*mut
f64} (i.e., \code{*double} in C), the developer should check that the storage
mode of \code{x} is indeed \samp{double} by checking against
\code{Rf\_isReal(x)}.  If not, a long jump will occur when calling
\code{REAL(x)}.

Care must also be taken when calling R API functions that might catch a user
interrupt (e.g, pressing \code{Ctrl-C} or hitting the stop button in RStudio)
because an interrupt also produces a long jump and leaks memory.  One R API
function that catches interrupts, for example, is the \code{Rprintf} function
for printing to R's console.

\section{High-level interface wrapping R's API}

To avoid the pitfalls of R API functions throwing errors or catching interrupts
when called from Rust, the \pkg{cargo} package also provides a high-level
interface defined in the \code{r} module.  This high-level
interface also alleviates the developer from deciding when results from R API
functions should be protected from the R's garbage collection and the necessary
bookkeeping involved in calling the \code{Rf\_unprotect} function.  Finally, the
high-level interface provides a more idiomatic API for Rust developers.  The
high-level interface is not a comprehensive wrapper over R's API, but it covers
common use cases and the developer can easily expand it by adding to the
\file{src/rustlib/roxido/src/r.rs} in the package.  That is, the developer does
not need to wait for the release of a new version of the \pkg{cargo} package.
The high-level interface provides a \code{check\_user\_interrupt} function to
test whether the user has tried to interrupt execution.  The \code{rprintln!}
macro behaves just like Rust's standard \code{println!} macro, but prints to the
R console and returns \code{true} if the user interrupted.  Much of the
interface is provided by associated functions for the \code{Rval} structure.
Recall that the
API can be browsed by executing \samp{cargo doc --open} when in the
\file{src/rustlib/roxido} directory.

The package generated by the \code{cargo::new\_package} function provides two
examples of the high-level interface.  These are translations of examples in
``Writing R Extensions''.  Consider first the \code{convolve2} function from
Section 5.10.1 ``Calling .Call''.  The translation is provided in
\file{src/rustlib/src/lib.rs} and shown below.
\begin{example}[numbers=left,firstnumber=21,xleftmargin=6.2mm]
    #[roxido]
    fn convolve2(a: Rval, b: Rval) -> Rval {
        let (a, xa) = a.coerce_double(&mut pc).unwrap();
        let (b, xb) = b.coerce_double(&mut pc).unwrap();
        let (ab, xab) = Rval::new_vector_double(a.len() + b.len() - 1, &mut pc);
        for xabi in xab.iter_mut() { *xabi = 0.0 }
        for (i, xai) in xa.iter().enumerate() {
            for (j, xbj) in xb.iter().enumerate() {
                xab[i + j] += xai * xbj;
            }
        }
        ab
    }
\end{example}
Notice on lines 23 and 24 the calls to \code{Rval}'s \code{coerce\_double} method.
The developer is encouraged to read the definition of this method in
\file{src/rustlib/roxido/src/r.rs}, but the gist of the method is to check R's
type of the \code{Rval} and convert it to R's storage mode \samp{double}, if
needed and if possible.  The method returns either a tuple giving a
(potentially-new) \code{Rval} and an \code{f64} slice into it, or an error.  If
the developer is confident that the  method will not fail, the developer can
simply call the \code{unwrap} method, as in lines 23 and 44, but more formal error
handling can be implemented in the usual Rust manner.  If \code{unwrap} is
called on an error message, the code will panic and a helpful message regarding
the location of the panic is displayed in the R console.  No memory leak occurs
and the R session is still valid. Thus, panics in the \pkg{cargo} framework are
controlled events.

In contrast to the \code{coerce\_double} method, a slice into R's memory for
vectors of doubles, integers, and logicals can be obtained without a potential
memory allocation using \code{x.slice\_double()}, \code{x.slice\_integer()}, and
\code{x.slice\_logical()} when \code{x} is an \code{Rval}.  In any case, these
slices are views into R's internal memory. Care should be taken when dealing
with R's special values.  For example, R's \code{NA} value for an element of an
\samp{integer} vector corresponds to Rust's \code{i32::MIN} (which is not a
special value in Rust). So, for example, \code{NA\_integer\_ * 0L} in R equals
\code{NA\_integer\_}, but it equals \code{0} in Rust.  Associated functions,
such as \code{Rval::is\_na\_integer}, are provided to test against R's special
values.  See Section 5.10.3 ``Missing and special values'' in ``Writing R
Extensions'' for a discussion of this issue.

Notice the argument to the \code{coerce\_double} method on lines 23 and 24 is
\code{\&mut pc}. The wrapper code provided by the \code{\#[roxido]} attribute
includes \code{let mut pc = Pc::new()}.  Many of the functions take a shared
mutable reference to a \code{Pc} structure.  The purpose of the \code{Pc}
structure is to handle the bookkeeping associated with \code{Rf\_protect} and
\code{Rf\_unprotect} calls related to R's garbage collection. It has a single public
method \code{protect} which takes an \code{SEXP}, calls \code{Rf\_protect} on
it, increments an interval counter, and returns the \code{SEXP}. When an
instance of the \code{Pc} structure goes out of scope, the Rust compiler
automatically inserts a call to its associated \code{drop} function which calls
\code{Rf\_unprotect(x)} using its interval counter \code{x}.  Not only does the
developer not need to manually track the number of protected items, the
developer does not need to worry about when a value should be protected. If the
method requires a shared mutable reference to a \code{Pc}, then protection is
needed and automatically handled by the function and not the developer.

Now consider the \code{zero} function described in Section 5.11.1
``Zero-finding'' of ``Writing R Extensions''.  The translation to the
\pkg{cargo} framework is provided in the package generated by the
\code{new\_package} function. The code is provided in
\file{src/rustlib/src/lib.rs} and shown below. As with the previous \code{convolve2}
function, this is a ``drop-in'' replacement for the function defined in
``Writing R Extensions''.
\begin{example}[numbers=left,firstnumber=35,xleftmargin=6.2mm]
    #[roxido]
    fn zero(f: Rval, guesses: Rval, stol: Rval, rho: Rval) -> Rval {
        let slice = guesses.slice_double().unwrap();
        let (mut x0, mut x1, tol) = (slice[0], slice[1], stol.as_f64());
        if tol <= 0.0 { panic!("non-positive tol value"); }
        let symbol = Rval::new_symbol("x", &mut pc);
        let feval = |x: f64| {
            let mut pc = Pc::new();
            symbol.assign(Rval::new(x, &mut pc), rho);
            f.eval(rho, &mut pc).unwrap().as_f64()
        };
        let mut f0 = feval(x0);
        if f0 == 0.0 { return Rval::new(x0, &mut pc); }
        let f1 = feval(x1);
        if f1 == 0.0 { return Rval::new(x1, &mut pc); }
        if f0 * f1 > 0.0 { panic!("x[0] and x[1] have the same sign"); }
        loop {
            let xc = 0.5 * (x0 + x1);
            if (x0 - x1).abs() < tol { return Rval::new(xc, &mut pc); }
            let fc = feval(xc);
            if fc == 0.0 { return Rval::new(xc, &mut pc); }
            if f0 * fc > 0.0 { x0 = xc; f0 = fc; } else { x1 = xc; }
        }
    }
\end{example}
This example shows the creation of new R objects from Rust values (e.g., lines
40, 43, 47, etc.) and extracting Rust values from R objects (e.g., 37, 38, and 44).
Line 44 demonstrates evaluating an R expression such that errors are caught
rather than causing a long jump.  Again, the full high-level API can be browsed
by executing \samp{cargo doc --open} when in the \file{src/rustlib/roxido}
directory.

\section{Miscellaneous: Threading issues and seeding a RNG}

Rust supports ``fearless concurrency,'' making it safe and easy to harness the
power of multiple CPU cores.  One should bear in mind, however, that R's
internals are fundamentally designed for single-threaded access.  Any callbacks
into R (using the low-level or high-level interface) should come from the same
thread from which R originally called the Rust code.

R users expect to get reproducible results from simulation code when they use
R's \code{set.seed} function.  There are two options for Rust code to achieve
this: (i) produce random numbers using R's API (as in the previous
\code{myrnorm} example) or (ii) seed a Rust random number generator from R's
random number generator.  To aid with the second approach, the \pkg{cargo}
framework provides the \code{random\_bytes} function.

\section{Embedding Rust code in an R script}

Beyond package development, the \pkg{cargo} package also supports defining
functions by embedding Rust code directly in an R script.  This facilitates
experimentation and avoids the need to set up a new R package. The approach,
however, loses the developer aids of an integrated development environment. As
such, it is only recommended to use this for small code snippets. To
demonstrate, consider the balanced linear assignment problem, a combinatorial
optimization problem in which $N$ workers are assigned to $N$ tasks such that
the sum of costs of getting all tasks completed is minimized. Suppose there are
four workers and tasks and the cost matrix in R is as follows, with each row being
the costs of the four tasks for a particular worker.
\begin{example}
        cost_matrix <- matrix(c(
            5, 9, 4, 6,
            8, 7, 8, 6,
            6, 7, 9, 3,
            2, 3, 3, 1
        ), nrow=4, byrow=TRUE)
\end{example}
The Hungarian algorithm \citep{kuhn1955hungarian} solves the linear assignment
problem and is implemented in the \CRANpkg{RcppHungarian} package
\citep{rcpphungarian} on CRAN.  The Jonker-Volgenant algorithm
\citep{jonker1987shortest}, however, is faster and available in the lapjv Rust
crate \citep{pkg.lapjv}.  The following code uses the \code{rust\_fn} from the
\pkg{cargo} package to define an R function based on embedded Rust code
utilizing the lapjv crate.
\begin{example}
    library("cargo")
    lapjv <- rust_fn(weights, dependencies='lapjv = "0.2.0"', '
        if !weights.is_square_matrix() || !weights.is_double_or_integer() {
            panic!("The weights argument must be a square numeric matrix.");
        }
        let weights_vec = weights.coerce_double(&mut pc).unwrap().1.to_vec();
        let n = weights.nrow();
        let weights = lapjv::Matrix::from_shape_vec((n, n), weights_vec).unwrap();
        let solution = lapjv::lapjv(&weights).unwrap().0;
        let cost = lapjv::cost(&weights, &solution[..]);
        let (pairs, slice) = Rval::new_matrix_integer(n, 2, &mut pc);
        for (i, x) in slice[..n].iter_mut().enumerate() { *x = i as i32 + 1; }
        let s = &mut slice[n..];
        for (i, y) in solution.into_iter().enumerate() { s[y] = i as i32 + 1; }
        let result = Rval::new_list(2, &mut pc);
        result.names_gets(Rval::new(["cost", "pairs"], &mut pc));
        result.set_list_element(0, Rval::new(cost, &mut pc));
        result.set_list_element(1, pairs);
        result
    ')
    lapjv(cost_matrix)
\end{example}
The \code{lapjv} function takes one unnamed argument \code{weights}, which is
passed to the embedded Rust code as the variable \code{weights} of type
\code{Rval}.  This code depends on version 0.2.0 of the lapjv crate and is
automatically downloaded and compiled by Cargo because of the argument
\code{dependencies='lapjv = "0.2.0"'} in the call to the \code{rust\_fn}
function.  Downloading and compiling the dependencies can take several seconds,
but subsequent compilations are very fast due to caching. For example, on our
machine, the first compilation took 12.95 CPU seconds and 6.57 elapsed seconds,
whereas recompilation of slightly changed code only took 1.81 CPU seconds and
0.97 elapsed seconds.  This caching persists between R sessions.
When a function defined by \code{rust\_fn} is garbage collected, its associated
shared library is automatically unloaded.

Running the code produces a list giving the total cost and a matrix which pairs
each worker to a task.  Note that when the cost matrix is 1000 $\times$ 1000
of standard normal values, this implementation takes only 0.068 seconds whereas
the \pkg{RcppHungarian} package finds the same solution in 4.866 seconds, i.e., 70 times slower. The
point is not that C++ is slower than Rust, rather that the choice of
algorithms can be important and that the \pkg{cargo} package makes it easy to
pull in high quality Rust code from others with little effort.

\section{Benchmarks}

Here the overhead of calling a Rust function from R using our \pkg{cargo}
framework is investigated.  Benchmarks are shown against the \pkg{rextendr}
framework and the standard mechanism for calling a C function from R. For this
benchmark, we use version 0.1.37 of \pkg{cargo} and version 0.2.0 of
\pkg{rextendr}, running in R version 4.1.0. An algorithm that executes quickly
is purposefully used to benchmark the overhead of calling into and returning
from compiled code. Rust and C themselves are not benchmarked here, but the
reader is referred to The Computer Language Benchmarks Game
(\url{https://benchmarksgame-team.pages.debian.net/benchmarksgame/}), which
shows Rust beating GCC's C in about half of the benchmarks.

Consider various implementations to compute the Euclidean norm
\code{sqrt(sum(x\string^2))}.  Several versions based on \pkg{rextendr} are
provided to account for the following: (i) \pkg{rextendr} can automatically
convert \code{Robj} (its wrapper over an \code{SEXP} value) and many Rust types, and
(ii) when defining embedded functions, \pkg{rextendr} does not cache the lookup
of the function pointer.
\begin{example}
    writeLines(con="f_C.c", "
        #include <Rinternals.h>
        SEXP f_C(SEXP x) {
            int n = Rf_length(x); double *y = REAL(x); double ss = 0.0;
            for ( int i=0; i<n; i++ ) ss += y[i];
            return Rf_ScalarReal(sqrt(ss));
        }")
    system("R CMD SHLIB f_C.c")
    dyn.load("f_C.so")
    .f_C <- getNativeSymbolInfo("f_C", "f_C")$address
    f_C <- function(x) .Call(.f_C, x)

    f_cargo <- cargo::rust_fn(x, '
        let ss = x.slice_double().unwrap().iter().fold(0.0, |s,z| s + (*z)*(*z));
        Rval::new(ss.sqrt(), &mut pc)
    ')

    rextendr::rust_function('fn f_rextendr1(x: Robj) -> Robj {
        let ss = x.as_real_slice().unwrap().iter().fold(0.0, |s,z| s + (*z)*(*z));
        Robj::from(ss.sqrt())
    }')

    rextendr::rust_function('fn f_rextendr2(x: &[f64]) -> f64 {
        let ss = x.iter().fold(0.0, |s,z| s + (*z)*(*z));
        ss.sqrt()
    }')

    .f1 <- getNativeSymbolInfo("wrap__f_rextendr1", "librextendr1")$address
    f_rextendr1_cached <- function(x) .Call(.f1, x)

    .f2 <- getNativeSymbolInfo("wrap__f_rextendr2", "librextendr2")$address
    f_rextendr2_cached <- function(x) .Call(.f2, x)

    x <- rnorm(10)
    microbenchmark::microbenchmark(f_C(x), f_cargo(x), f_rextendr1(x), f_rextendr2(x),
        f_rextendr1_cached(x), f_rextendr2_cached(x), times=1000000)
\end{example}
A summary of the performance is included below.  Notice that the implementation
based on \pkg{cargo} is competitive with the C version and faster than the
\pkg{rextendr} implementations.
\begin{example}
        Unit: nanoseconds
                          expr  min   lq      mean median   uq      max neval
                        f_C(x)  349  504  616.3403    545  597  2329566 1e+06
                    f_cargo(x)  389  523  684.6221    563  625 54627019 1e+06
                f_rextendr1(x) 4918 5818 6282.8506   6011 6226  3925556 1e+06
                f_rextendr2(x) 3789 4504 4941.8917   4707 4918  6396105 1e+06
         f_rextendr1_cached(x) 4244 5197 5606.6939   5391 5586  6669919 1e+06
         f_rextendr2_cached(x) 3145 3849 4256.3246   4070 4278  6602903 1e+06
\end{example}

\section{Summary}

The hope is that this paper contributes to interest in developing R packages
with Rust.  The paper highlights idiosyncrasies of R and Rust that must be
addressed by any integration. The \pkg{cargo} framework provides a Rust
interface for commonly used parts of the R API that can easily be extended to
cover other parts of the R API. The framework minimizes the runtime overhead and
seeks to be transparent on how it interfaces R and Rust.

\bibliography{dahl}

\address{David B.\ Dahl\\
  Brigham Young University\\
  Provo, Utah\\
  University States of America\\
  (ORCiD: 0000-0002-8173-1547)\\
  \email{dahl@stat.byu.edu}}

\end{article}

\end{document}